\documentclass[lettersize,journal]{IEEEtran}
\usepackage{amsmath,amsfonts}
\usepackage{algorithmic}
\usepackage{algorithm}
\usepackage{array}
\usepackage[caption=false,font=normalsize,labelfont=sf,textfont=sf]{subfig}
\usepackage{textcomp}
\usepackage{stfloats}
\usepackage{xcolor}
\usepackage{multirow}
\usepackage{longtable}
\usepackage{tabularx}
\usepackage{url}
\usepackage{verbatim}
\usepackage{graphicx}
\usepackage{cite}
\usepackage{comment}
\begin{document}

\title{ VR Based Emotion Recognition Using Deep Multimodal Fusion With  Biosignals Across Multiple Anatomical Domains}

\author{Pubudu L. Indrasiri,~\IEEEmembership{Member,~IEEE}, Bipasha Kashyap,~\IEEEmembership{Member,~IEEE}, Chandima Kolambahewage,~\IEEEmembership{Member,~IEEE}, Bahareh Nakisa, Kiran Ijaz, Pubudu N. Pathirana,~\IEEEmembership{Senior Member,~IEEE}
\thanks{Pubudu L. Indrasiri and Bipasha Kashyap and Pubudu N. Pathirana are with the School of Engineering, Deakin University, Waurn Ponds, VIC 3216, Australia (e-mails: \{pranpatidewage, b.kashyap, pubudu.pathirana\}@deakin.edu.au).}

\thanks{Bahareh Nakisa is with the School of Information Technology, Faculty of Science Engineering and Built Environment, Deakin University, Burwood, VIC 3125, Australia (e-mail: bahar.nakisa@deakin.edu)}
\thanks{Kiran Ijaz is with the Wellbeing-supportive Technology Laboratory,
School of Electrical and Information Engineering, The University of Sydney, NSW 2006 Australi (e-mail: kiran.ijaz@sydney.edu.au)}
\thanks{Chandima Kolambahewage is with the School of Engineering, Faculty of
Science Engineering and Built Environment, Deakin University, Waurn Ponds,
VIC 3216, Australia (e-mail: c.kolambahewage@deakin.edu)}

}

\markboth{Journal of \LaTeX\ Class Files,~Vol.~14, No.~8, August~2021}%
{Shell \MakeLowercase{\textit{et al.}}: A Sample Article Using IEEEtran.cls for IEEE Journals}


\maketitle

\begin{abstract}
Emotion recognition is significantly enhanced by integrating multimodal biosignals and IMU data from multiple domains. In this paper, we introduce a novel multi-scale attention-based LSTM architecture, combined with Squeeze-and-Excitation (SE) blocks, by leveraging multi-domain signals from the head (Meta Quest Pro VR headset), trunk (Equivital Vest), and peripheral (Empatica Embrace Plus) during affect elicitation via visual stimuli. Signals from 23 participants were recorded, alongside self-assessed valence and arousal ratings after each stimulus. LSTM layers extract features from each modality, while multi-scale attention captures fine-grained temporal dependencies, and SE blocks recalibrate feature importance prior to classification. We assess which domain's signals carry the most distinctive emotional information during VR experiences, identifying key biosignals contributing to emotion detection. The proposed architecture, validated in a user study, demonstrates superior performance in classifying valance and arousal level (high / low), showcasing the efficacy of multi-domain and multi-modal fusion with biosignals (e.g., TEMP, EDA) with IMU data (e.g., accelerometer)  for emotion recognition in real-world applications.
\end{abstract}



\section{Introduction}
\IEEEPARstart{E}{motion} recognition has emerged as a critical research area in recent years, driven by advancements in deep learning and the increasing availability of multimodal data sources \cite{yang2021behavioral}, \cite{katsigiannis2017dreamer}, \cite{koelstra2011deap}, \cite{zhang2020emotion}. Affect recognition is central to various applications in human-computer interaction (HCI), healthcare, gaming, and adaptive systems, where an accurate understanding of emotional states can significantly enhance user experience and system responsiveness. Traditionally, emotion detection has relied on unimodal data streams, such as physiological signals, motion data, or facial expressions \cite{pan2023review}. However, human emotions' inherent complexity and dynamic nature necessitate more sophisticated approaches, prompting a shift toward multimodal systems that integrate data from multiple sources to improve classification accuracy and robustness.

\par Physiological signals—such as electrocardiograms (ECG), electrodermal activity (EDA), and heart rate (HR)—have demonstrated strong correlations with emotional arousal, stress, and other affective states  \cite{katsigiannis2017dreamer}, \cite{fan2023new}, \cite{yang2021behavioral} , \cite{song2019mped}. Similarly, motion data captured by inertial measurement units (IMUs) has been employed to infer emotions through analysis of body movements \cite{hashmi2020motion}, \cite{ma2020smart}. Eye-tracking, which provides insight into gaze behavior and visual focus, further contributes contextual understanding of emotional states \cite{alghowinem2014exploring}, \cite{mou2021driver}. Despite the strengths of these individual modalities, each has inherent limitations when used in isolation. For example, while physiological signals capture autonomic nervous system activity, they do not account for contextual or environmental factors that may be captured through motion data or gaze tracking.

\par To address these limitations, multimodal data integration has gained significant attention \cite{koelstra2011deap}, \cite{mou2021driver}, \cite{yang2021behavioral}. Multimodal systems leverage the complementary strengths of different data streams, providing a more comprehensive understanding of emotional states. Recent advances in deep learning have enabled the efficient fusion of high-dimensional data from multiple modalities, offering significant improvements over traditional machine learning techniques in feature extraction and pattern recognition.

\par In this work, we propose a novel emotion recognition framework that integrates data from three distinct devices, each corresponding to a different physiological and motion domain: the trunk, head, and peripheral. The devices employed include a wearable vest (trunk domain), a virtual reality (VR) headset (head domain), and a wrist-worn device (peripheral domain). Each device contributes unique data modalities—physiological signals (e.g., ECG and HR) from the vest, motion data from the wrist via accelerometers, and feature-extracted data from the VR headset. By incorporating these multimodal inputs, we aim to capture emotional states with greater precision, leveraging the complementary nature of data across the trunk, head, and peripheral domains.

\par This multi-domain approach addresses several key challenges in affect recognition. By capturing physiological signals from the trunk, motion data from the wrist, and gaze behaviour from the head, we construct a more holistic view of emotional states, thus overcoming the limitations of single-modality systems. Additionally, the fusion of these modalities enables a more accurate classification of complex emotional responses, particularly in dynamic or interactive environments such as virtual reality.

\par This paper presents the methodology for collecting and processing data from three critical domains: peripheral, trunk, and head, and the development of a deep learning-based multimodal architecture for valence and arousal detection. Our proposed architecture integrates Long Short-Term Memory (LSTM) networks with multi-scale attention mechanisms for feature extraction from unimodal signals within each domain. Domain-wise fusion is then applied, followed by Squeeze-and-Excitation (SE) blocks to dynamically recalibrate domain outputs. The recalibrated outputs from the three domains are subsequently fused into a multimodal vector, which is passed to the output layer for emotion classification. We provide comprehensive experimental results that highlight the effectiveness of this multi-domain, multimodal approach, showcasing its superiority over unimodal systems in terms of both accuracy and robustness, with substantial performance gains in emotion detection.

\section{Related Works}
In recent years, affect recognition has garnered significant attention, particularly with the rise of multimodal data sources and advanced deep learning techniques \cite{khan2022evaluation}, \cite{sarkar2020self}, \cite{siriwardhana2020multimodal}. Numerous studies have explored various approaches for recognizing affective states using physiological signals and additional modalities such as motion or environmental sensors.

\subsection{Emotion Detection Using Physiological Signals}
Physiological signals have been extensively explored in the field of affective computing for emotion detection, given their strong correlation with emotional states. Among the most commonly utilized signals is the electrocardiogram (ECG), which provides insights into heart rate variability (HRV), a key indicator of autonomic nervous system activity and emotional arousal \cite{katsigiannis2017dreamer}, \cite{fan2023new},\cite{song2019mped}. Electrodermal activity (EDA), also referred to as galvanic skin response (GSR), measures skin conductance and reflects sympathetic nervous system responses to emotional stimuli, making it a vital tool for detecting arousal and stress \cite{yang2021behavioral}, \cite{song2019mped}. Additionally, blood volume pulse (BVP), typically measured via photoplethysmography (PPG), has been widely used to assess heart rate and peripheral blood flow, contributing to the detection of emotional responses related to stress and anxiety \cite{yang2021behavioral}. Other crucial signals include respiration rate (RR), which varies with emotional states such as fear or relaxation, and electromyography (EMG), which captures facial muscle movements linked to expressions of emotion \cite{sarkar2021simultaneous}. Moreover, electroencephalogram (EEG) has been extensively employed to monitor brainwave activity, offering insights into the valence and arousal dimensions of emotion \cite{katsigiannis2017dreamer} ,\cite{song2019mped}, \cite{koelstra2011deap}, \cite{nakisa2018evolutionary}. Finally, skin temperature (ST) and heart rate (HR) have also proven useful in detecting stress and emotional arousal, further contributing to a holistic understanding of affective states \cite{yang2021behavioral}. These signals, often used in combination, provide a rich, multimodal approach to accurately capturing and classifying complex emotional experiences.

\subsection{Emotion Detection Using Motion Signals}
Emotion recognition research based on human motion has attracted significant attention within the field of human-computer interaction, particularly with the advancements in IMU sensors, which have facilitated efficient and seamless interaction between humans and devices \cite{hashmi2020motion}, \cite{ma2020smart}, \cite{leng2024emotion}, \cite{rahmani2024emowear}. IMU sensors, such as accelerometers (ACC) and gyroscopes (GYRO), capture crucial motion-related data that can be used to infer emotional states. Recent studies have explored the use of wearable inertial devices attached to various body parts to collect emotional information through human movement. Hashmi et al. \cite{hashmi2020motion} employed a smartphone worn on the chest to record human gait signals and used spectro-temporal features from stride signals for emotion recognition, identifying six basic emotions using Random Forest and SVM classifiers. Similarly, Gravina et al. \cite{ma2020smart} leveraged sensor-level and feature-level fusion to monitor in-seat activities, which reflect psychological and emotional states, though the handcrafted features required labour-intensive design. Chang et al. \cite{chang2022retracted} used intelligent inertial sensors to recognize emotions during badminton play, though prolonged activity resulted in physical fatigue that affected the accuracy of emotion detection. Furthermore, recent advancements such as those proposed by Feng et al. \cite{yu2024intelligent} have integrated motion and emotion recognition into intelligent wearable systems using multi-sensor fusion, enabling real-time interaction and feedback in virtual environments via Digital Twin technology, significantly enhancing user experience and emotion recognition accuracy. These studies highlight the increasing potential of IMU-based motion signals in emotion recognition, though challenges such as feature extraction complexity and physical fatigue remain areas of ongoing research.

\subsection{Emotion Detection Using Eye Tracking Data}
Humans interact with their environment, with each elicited emotion being closely tied to its specific context \cite{valtchanov2015enviropulse}. Ptaszynski et al. \cite{ptaszynski2009towards} highlighted the importance of incorporating contextual analysis into emotion processing. Eye-tracking involves identifying a user’s gaze point or focus on a specific visual stimulus. An eye-tracker, a device designed for this purpose, measures eye position and movements \cite{strandvall2009eye}. As a sensor technology, eye-tracking is versatile, applicable in diverse configurations and uses, as demonstrated by Singh et al. \cite{singh2012human}. Furthermore, other studies explore eye movements as potential indicators for recognizing emotions \cite{alghowinem2014exploring}, \cite{mou2021driver}.

\subsection{Multimodal Emotion Detection}
\label{model fusion}


Multimodal fusion has captured significant interest across various research domains due to its broad applicability in areas such as emotion recognition, event detection, image segmentation, and video classification \cite{lahat2015multimodal}. Fusion strategies are traditionally categorized by the fusion level, including: 1) feature-level fusion (early fusion), 2) decision-level fusion (late fusion), and 3) hybrid multimodal fusion. With advancements in deep learning, an increasing number of researchers are leveraging deep learning frameworks to enhance multimodal fusion approaches.

Feature-level fusion is a widely adopted approach for integrating multiple modalities, where features from each modality are combined into a high-dimensional representation and then processed as a unified input by models \cite{lu2015combining}, \cite{koelstra2011deap}, \cite{mou2021driver}, \cite{yang2021behavioral}.

Decision-level fusion, by contrast, involves the use of multiple classifiers and their aggregation, frequently through ensemble learning techniques \cite{zhou2012ensemble}. This approach merges individual classifier outputs into a single decision. Techniques such as MAX fusion, SUM fusion, and fuzzy integral fusion are commonly employed in multimodal emotion recognition, highlighting the complementary nature of EEG and eye movement features through confusion matrix analysis \cite{liu2021comparing}, \cite{song2018decision}.

\subsection{Emotion Detection Using Multi-Domain Sensors}
Emotion recognition using IMU sensors and biosensors in VR environments has emerged as a critical area of research, with a significant focus on analyzing signals from isolated domains such as head, trunk, and peripheral sensors. Historically, emotion detection efforts have been domain-specific \cite{pan2023review}. For example, head-mounted sensors, predominantly integrated into VR headsets, capture signals like eye movement and facial EMG to infer emotional states. Trunk-mounted sensors have primarily focused on physiological signals such as HRV and respiration, while peripheral devices, particularly wrist-worn sensors, monitor EDA, skin temperature, and accelerometry. While these single-domain approaches have demonstrated effectiveness in capturing emotional cues, they inherently limit the understanding of how emotions are represented across the body in different regions \cite{yang2021behavioral},\cite{pan2023review}.

However, a critical gap in the current literature is the lack of comparative studies that examine the relative effectiveness of these domains (head, trunk, and peripheral) in emotion recognition tasks. Specifically, no substantial research has been conducted to systematically compare the emotional information captured by sensors in these distinct body regions within the same experimental setup. Furthermore, the literature has not adequately explored the potential benefits of multimodal sensor fusion, where signals from multiple domains are combined to exploit complementary features and improve emotion detection accuracy. This omission leaves a gap in understanding the holistic contribution of multi-domain signals in representing complex emotional states, particularly in immersive VR settings.

\subsection{Motivations and Contributions}

Building on the identified gaps in single-domain emotion recognition, this study aims to leverage the potential of multi-domain sensor fusion in VR environments. By systematically comparing signals from the head, trunk, and peripheral regions, we seek to determine whether the combination of signals from these diverse domains can provide a more comprehensive and accurate representation of emotional states. Additionally, this work explores the specific contributions of each domain and evaluates the benefits of integrating these signals for more precise emotion detection.

This study advances multi-domain inference in VR emotion recognition by presenting a framework, within the same experimental setup, that:

\begin{itemize}
\item 
Identifies Key Emotional Domains:
Although emotion originates from the central nervous system, emerging research supports the idea that physiological signals from various body domains such as the head \cite{alghowinem2014exploring}, \cite{mou2021driver}, trunk \cite{katsigiannis2017dreamer}, \cite{fan2023new},\cite{song2019mped}, and peripheral \cite{8599278}, \cite{chang2022retracted} systems, offer unique and complementary emotional insights. When these signals are integrated, they significantly enhance the accuracy and robustness of emotion recognition systems. Our comprehensive analysis of these domains within the same experimental setup elucidates which signals contribute most effectively to emotional inference during VR experiences, ultimately advancing the precision of emotion recognition in VR environments. 
\item Performs Biosignal Importance Analysis: Using advanced techniques such as deep learning, we pinpoint specific biosignals within the most informative domain that contribute most significantly to emotion recognition, enhancing model interpretability and optimization.

\item Propose a novel deep learning-based multi-modal architecture that leverages multi-scale attention mechanisms and Squeeze-and-Excitation (SE) blocks. It captures fine-grained dependencies across modalities and dynamically recalibrates feature representations to enhance accuracy and robustness in emotion detection.

\item Enables Multimodal Data Fusion: We assess the performance of a fused multi-domain model to evaluate whether combining data across domains yields superior accuracy for emotion detection compared to single-domain approaches.
\end{itemize}

This multi-domain framework provides a scalable and precise tool for VR emotion recognition research, with implications for enhanced affective computing, user experience analysis, and behavioural studies in immersive environments.


\section{Experimental Protocol}

\subsection{Video Stimuli Selection}
\begin{table*}[h]
\caption{Comprehensive list of all immersive VR clips used in this experiment. For the first 8 VR clips, the respective quadrants were explicitly provided in the literature \cite{voigt2020comparing}. For the remaining 5 VR clips, quadrant assignments were not given \cite{guimard2022pem360}; instead, ratings on valence and arousal scales were provided. Ratings from 1-4 were categorized as low (LV for valence, LA for arousal), and ratings from 4-7 were classified as high (HV for valence, HA for arousal). The majority label from participants’ responses was selected as each video's final label and the majority percentage is given for each video's valence and arousal as shown in the SAM Questionnaire (Majority) column.}
\resizebox{\textwidth}{!}{%
\begin{tabular}{lllllllllll}
\hline
\multirow{2}{*}{\textbf{S.No.}} & \multirow{2}{*}{\textbf{Title (ID)}} & \multirow{2}{*}{\textbf{Description}} & \multirow{2}{*}{\textbf{Duration}} & \multirow{2}{*}{\textbf{Start}} & \multirow{2}{*}{\textbf{End}} & \multicolumn{3}{c}{\textbf{Literature}} & \multicolumn{2}{c}{\textbf{SAM Questionnaire (Majority)}} \\ \cline{7-11} 
 &  &  &  &  &  & \textbf{Arousal (G2\_A)} & \textbf{Valence (G2\_V)} & \textbf{Quadrant} & \textbf{Arousal (G1\_A)} & \textbf{Valence (G1\_V)} \\ \hline
1 & Jailbreak (68) & Viewer takes the perspective of a fighter jet pilot in an incarcerated air hangar. & 60 & 2:39 & 3:39 & HA & LV & HALV & HA 82.61\% & LV 73.91\% \\ \hline
2 & War knows no nation (20) & Short film where a group of soldiers defend against a zombie attack. & 60 & 4:44 & 5:44 & HA & LV & HALV & HA 82.61\% & LV 73.91\% \\ \hline
3 & The displaced (18) & Short film on the emotional trauma of solitary confinement. & 60 & 2:23 & 3:23 & LA & LV & LALV & HA 65.22\% & LV 91.30\% \\ \hline
4 & Solitary confinement (16) & Short film on the emotional trauma of solitary confinement. & 60 & 0:00 & 1:00 & LA & LV & LALV & HA 65.22\% & LV 91.30\% \\ \hline
5 & Walk the tight rope (69) & Viewer experiences walking a tight rope over a canyon. & 60 & 0:27 & 1:27 & HA & HV & HAHV & LA 52.17\% & HV 86.96\%\\ \hline
6 & Puppies host SourceFed for a day (50) & Show puppies host the SourceFed team for a day. & 60 & 0:04 & 1:04 & HA & HV & HAHV & LA 52.17\% & HV 86.96\% \\ \hline
7 & Malaekahana Sunrise (32) & Time-lapse clip showing the sun rising over a forest. & 60 & 1:20 & 2:20 & LA & HV & LAHV & LA 60.87\% & HV 100\% \\ \hline
8 & Great ocean road (22) & Aerial shots over various scenic locations in Australia. & 60 & 0:00 & 1:00 & LA & HV & LAHV & LA 60.87\% & HV 100\% \\ \hline
9 & The fight to save threatened species (12) & Various shots of animals in captivity and on the brink of extinction. & 98 & 0:05 & 1:03 & HA (4.6) & HV (7) & - & LA 60.87\% & HV 86.96\% \\ \hline
10 & Instant Caribbean vacation (23) & Promotional video of a Caribbean cruise liner. & 135 & 0:09 & 1:10 & LA (3.2) & HV (7.2) & - & LA 52.17\% & HV 86.96\% \\ \hline
11 & Seagulls (27) & Viewer sees the sun rising over the horizon at a beach. & 120 & 1:00 & 2:20 & LA (1.6) & HV (6.0) & - & LA 73.91\% & LV 56.52\%\\ \hline
12 & The margins (13) & Various shots of animals in captivity and on the brink of extinction. & 127 & 0:04 & 1:31 & HA (4.08) & HV (4.92) & - & HA 56.52\% & LV 91.30\% \\ \hline
13 & Through Mowgli's Eyes (73) & \begin{tabular}[c]{@{}l@{}}Viewer takes the perspective of an extreme sports participant leaping \\ down a huge slope before leaping across an open expanse.\end{tabular} & 61 & 0:09 & 0:09 & HA (6.18) & HV (6.27) & - & HA 60.87\% & LV 52.17\% \\ \hline
\end{tabular}%
}\label{ref:video selection}
\end{table*}
Thirteen 360-degree videos were sourced from the public database described in \cite{li2017public}, specifically chosen for their ability to evoke targeted emotional responses. Due to the varied lengths of these original videos, three domain experts—two psychologists and one computer scientist—selected specific intervals within each video that best represented the intended emotional state for viewers \cite{guimard2022pem360,voigt2020comparing}. This standardization ensured uniform stimulus duration across all videos, reducing variability in the experimental setup and maintaining an efficient overall study duration.

\par Table I lists the videos used, detailing the selected intervals and associated emotional quadrants (LALV = low arousal, low valence; LAHV = low arousal, high valence; HALV = high arousal, low valence; HAHV = high arousal, high valence). Video clips were categorized by valence and arousal levels, grouping clips of similar emotional tones within the same quadrant. Participants were presented with a printed Self-Assessment Manikin (SAM) scale reference, ratings  $ < 4 $ as low valence (LV) and low arousal (LA), and ratings $ > 4 $ as high valence (HV) and high arousal (HA) and instructed to rate each video group on these dimensions after viewing, allowing for consistent assessment of emotional responses (Figure \ref{f07}).

\subsection{Participants}
In this study, a cohort of 23 healthy participants was recruited through digital and physical advertisements displayed across social media platforms and within the university campus. A substantial portion of the sample (75.0\%, n=18) comprised male participants, while the remaining 25.0\% (n=6) were female. The age of participants ranged from 21 to 52 years, yielding a mean age of 28.9 years with a standard deviation of 6.4 years. All participants were screened to confirm the absence of any pre-existing cardiovascular or psychological conditions, ensuring a baseline of physiological and psychological health across the sample.
\subsection{Experimental Setup}
\begin{figure*}
	\centering
	\includegraphics[width=.99\linewidth]{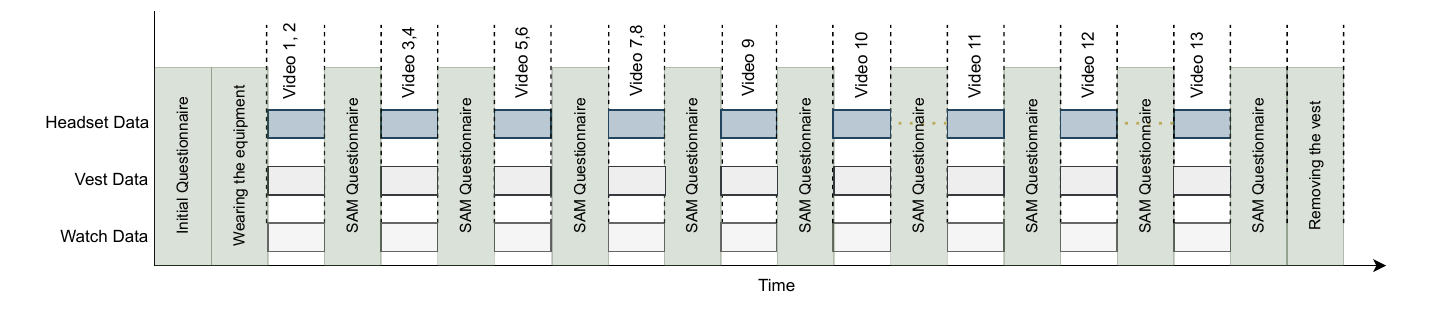}
	\caption{Data collection arrangement for a single participant.}
	\label{f07}
\end{figure*}

\begin{figure*}
	\centering
	\vspace{-0.1in}
 \includegraphics[width=.99\linewidth]{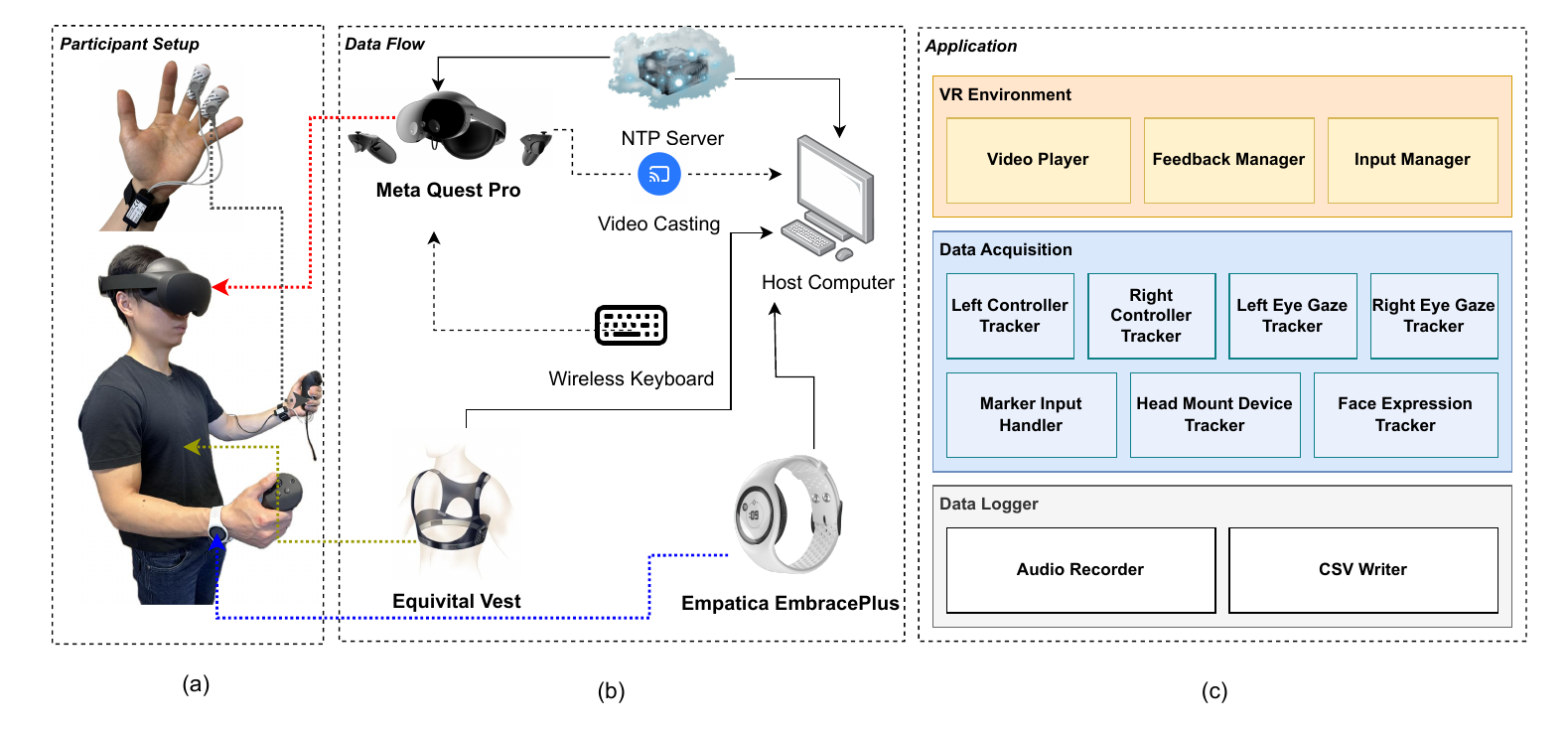}
	\caption{End-to-end data collection setup. (a) a participant wearing the Equivital Vest (inside his shirt), Empatica EmbracePlus and Meta Quest Pro VR headset with two controllers in his hands. Gel electrodes for measuring GSR activity are attached to the middle and ring fingers of the left hand. The participant is asked to comfortably move around and interact with the VR content while standing, (b) the high level data flow diagram across three domains, (c)  bespoke VR application architecture. }
	\label{data collection arc}
	\vspace{-0.1in}
\end{figure*}

Three commercially available devices—the Equivital Vest, Empatica Embrace Plus Watch, and Meta Quest Pro VR headset were employed to collect a range of physiological and behavioural measurements from participants. The Equivital Vest (specifically, \textit{Equivital EQ02+ LifeMonitor}) was utilized to capture electrocardiogram (ECG) data along with accelerometer readings, while the Empatica Watch (specifically, \textit{Empatica E4 Watch} recorded EDA, skin temperature, heart rate, and accelerometer data. The Equivital GSR add-on accessory was used to capture GSR (Galvanic Skin Response) activity.
 The \textit{Meta Quest Pro VR headset}, in addition to delivering immersive 3D 360-degree video stimuli, provided critical eye-tracking data, capturing participant gaze patterns and reactions in real-time. Within the same experimental set-up, sensory data from the trunk, peripheral (wrist), and head domains were captured using the \textit{Equivital Vest}, \textit{Empatica Embrace Plus Watch}, and \textit{Meta Quest Pro VR headset}, respectively. Figure \ref{data collection arc}.(b) provides a schematic overview of the data collection architecture.


To achieve temporal precision across all devices in the experimental setup, the \textit{Host Computer} and \textit{Meta Quest Pro} were synchronized via a shared Network Time Protocol (\textit{NTP}) server. This alignment ensured uniform timekeeping, critical for data accuracy. Additionally, the control applications for the \textit{Equivital Vest} and \textit{Empatica Watch}, running on the \textit{host computer}, upheld the same time synchronization protocol, reinforcing the cohesion across the system. Data recording on both the vest and watch commenced automatically upon activation and continued until device removal and shutdown. User control over the \textit{Meta Quest Pro} was enabled through a \textit{wireless keyboard}, while the research team monitored the participant's perspective in real-time through the device’s live video streaming capability. A custom-built VR \textit{video player} and \textit{data logger} application developed using Unity and C\#, facilitated seamless, real-time data capture from the headset’s integrated sensors. Figure \ref{data collection arc}.(c) illustrates the architecture of this custom VR application.

This multi-modal data collection framework, which incorporated physiological signals, immersive environments, and eye-tracking data, provided a robust platform for evaluating participant responses and behaviours under controlled conditions.

The VR application architecture is structured into three core modules to optimize immersive and interactive data collection: the \textit{VR Environment}, \textit{Data Acquisition}, and \textit{Data Logger} modules. The VR Environment serves as the primary user interface, offering real-time interaction capabilities for both participants and researchers. The \textit{Data Acquisition} module systematically gathers data from multiple headset sensors, while the \textit{Data Logger} archives this information efficiently, ensuring that data is available for post-experiment analysis.

The \textit{Video Player} component enables the playback of audiovisual stimuli by loading predesignated media files directly from the headset’s file system, delivering an immersive, synchronized audiovisual experience.

An \textit{Input Manager} module tracks inputs from a \textit{wireless keyboard}, allowing the researcher to control playback and data logging functions dynamically. Through predefined keyboard shortcuts, the researcher can navigate media clips, initiate play/pause, skip content, insert event markers, and terminate the application. The \textit{Feedback Manager} enhances usability by providing a responsive visual overlay, confirming command execution (e.g., a pause icon during a video pause).

Within the \textit{Data Acquisition module}, the system utilizes the Meta XR SDK to interface with various headset sensors. The \textit{Head Mount Device Tracker} provides real-time updates on the headset’s spatial position and orientation relative to an auto-calibrated reference point. \textit{Left} and \textit{right} Controller Trackers capture hand movements' positional and rotational data relative to the headset, while the \textit{Eye Gaze Trackers} record gaze orientation and fixation points, offering insights into visual attention patterns. The \textit{Marker Input Handler} logs event markers initiated via the keyboard in CSV format, streamlining data synchronization and subsequent analysis.

The \textit{Data Logger} module consists of two sub-components: the \textit{Audio Recorder} and \textit{CSV Writer}. The \textit{Audio Recorder} captures high-fidelity audio input from the headset’s microphones, saving it in .wav format. Simultaneously, the \textit{CSV Writer} logs all sensor data with precise timestamps in a structured .csv format, ensuring thorough temporal alignment of multimodal data.

This multi-layered architecture enables rigorous data acquisition, reliable interaction, and enhanced data integrity, establishing the application as a scalable and robust tool for immersive experimental research.

\subsection{Data Collection}
Upon arrival, participants were briefed on the study’s objectives and protocol. Researchers ensured accurate equipment fitting after providing informed consent and completing a demographic survey.

Participants watched thirteen videos, completing the SAM Manikin Questionnaire after each video to rate their emotional responses on a scale from 1 to 7 for both valence and arousal. This process of viewing videos and completing the questionnaire was repeated across all video sets. Researchers then assisted with equipment removal, ensuring standardized data collection for robust, synchronized multi-modal analysis. The data collection process is illustrated in Fig \ref{f07}.


\subsection{Data Analysis and Preprocessing}

The system accepts raw data from Empatica Embrace Plus Watch, Equivital Vest and Meta Quest Pro VR Head Set as inputs.Through the data preprocessing module, the heterogeneous inputs can be formatted into specific representations, which can be effectively used in the following feature extraction network module.

\begin{table}
\caption{The different modalities collected from three domains, the  peripheral (Empatica EmbracePlus), trunk (Equivital Vest), and head (Meta Quest Pro VR Headset), along with their full names and corresponding abbreviations and sampling frequencies (in brackets). Only the abbreviations are used throughout this paper.}
\resizebox{\columnwidth}{!}{%
\begin{tabular}{ccc}
\hline
\textbf{\begin{tabular}[c]{@{}c@{}}Peripheral \\ (Empatica EmpracePlus)\end{tabular}} & \textbf{\begin{tabular}[c]{@{}c@{}}Trunk \\ (Equivital Vest)\end{tabular}} & \textbf{\begin{tabular}[c]{@{}c@{}}Head \\ (Meta Quest Pro)\end{tabular}} \\ \hline
\begin{tabular}[c]{@{}c@{}}Accelerometer\_x \\ (ACC\_X, 64 Hz)\end{tabular} & \begin{tabular}[c]{@{}c@{}}Electrocardiogram Lead 1 \\ (ECG1, 256 Hz)\end{tabular} & \begin{tabular}[c]{@{}c@{}}LEyeposition\_x\\ (L\_EP\_X)\end{tabular} \\ \hline
\begin{tabular}[c]{@{}c@{}}Accelerometer\_y \\ (ACC\_Y, 64 Hz)\end{tabular} & \begin{tabular}[c]{@{}c@{}}Electrocardiogram Lead 2 \\ (ECG2, 256 Hz)\end{tabular} & \begin{tabular}[c]{@{}c@{}}LEyeposition\_y\\ (L\_EP\_Y)\end{tabular} \\ \hline
\begin{tabular}[c]{@{}c@{}}Accelerometer\_z \\ (ACC\_Z, 64 Hz)\end{tabular} & \begin{tabular}[c]{@{}c@{}}Lateral Accelerometer \\ (LAT\_ACC, 256 Hz)\end{tabular} & \begin{tabular}[c]{@{}c@{}}LEyeposition\_z \\ (L\_EP\_Z)\end{tabular} \\ \hline
\begin{tabular}[c]{@{}c@{}}Temperature \\ (TEMP)\end{tabular} & \begin{tabular}[c]{@{}c@{}}Longitudinal Accelerometer \\ (LONG\_ACC, 256 Hz)\end{tabular} & \begin{tabular}[c]{@{}c@{}}REyeposition\_x \\ (R\_EP\_X)\end{tabular} \\ \hline
\begin{tabular}[c]{@{}c@{}}Electrodermal Activity \\ (EDA, 4 Hz)\end{tabular} & \begin{tabular}[c]{@{}c@{}}Vertical Accelerometer \\ (VERT\_ACC, 256 Hz)\end{tabular} & \begin{tabular}[c]{@{}c@{}}REyeposition\_y \\ (R\_EP\_Y)\end{tabular} \\ \hline
\begin{tabular}[c]{@{}c@{}}Blood Volume Pulse \\ (BVP, 64 Hz)\end{tabular} & \begin{tabular}[c]{@{}c@{}}Galvanic Skin Response \\ (GSR, 256 Hz)\end{tabular} & \begin{tabular}[c]{@{}c@{}}REyeposition\_z\\ (R\_EP\_Z)\end{tabular} \\ \hline
\end{tabular}%
}
\label{ref:domain info}
\end{table}

The signals from the Empatica EmbracePlus and Equivital Vest can be categorized into two primary types: physiological signals and inertial measurement unit (IMU) data. Physiological signals encompass measurements such as heart rate (ECG), skin conductance (EDA), blood volume pulse (BVP), and skin temperature (TEMP). The ECG, or electrocardiogram, records the electrical activity of the heart over time, providing detailed information about the heart's rhythm and rate by detecting the depolarization and repolarization of cardiac muscles. Blood volume pressure (BVP), measured via a photoplethysmography (PPG) sensor embedded in the watch, reflects the pulse wave of the heart and the volume of blood flowing through vessels, offering a non-invasive insight into cardiovascular dynamics. EDA, or electrodermal activity, captures variations in the skin's electrical conductance, which is influenced by sweat gland activity and serves as a physiological marker for emotional and arousal states. This signal is monitored by applying a low-level current between two electrodes on the skin, providing valuable data about the body's electrodermal response to external stimuli. Skin temperature (TEMP) is another critical metric, measured using a thermopile infrared sensor, which helps monitor the body's thermoregulatory response. Vascular responses, such as vasodilation or vasoconstriction, result in corresponding changes in skin temperature, which can indicate the body's reaction to various environmental and physiological factors.
\begin{figure*}
	\centering
	\includegraphics[width=1\linewidth]{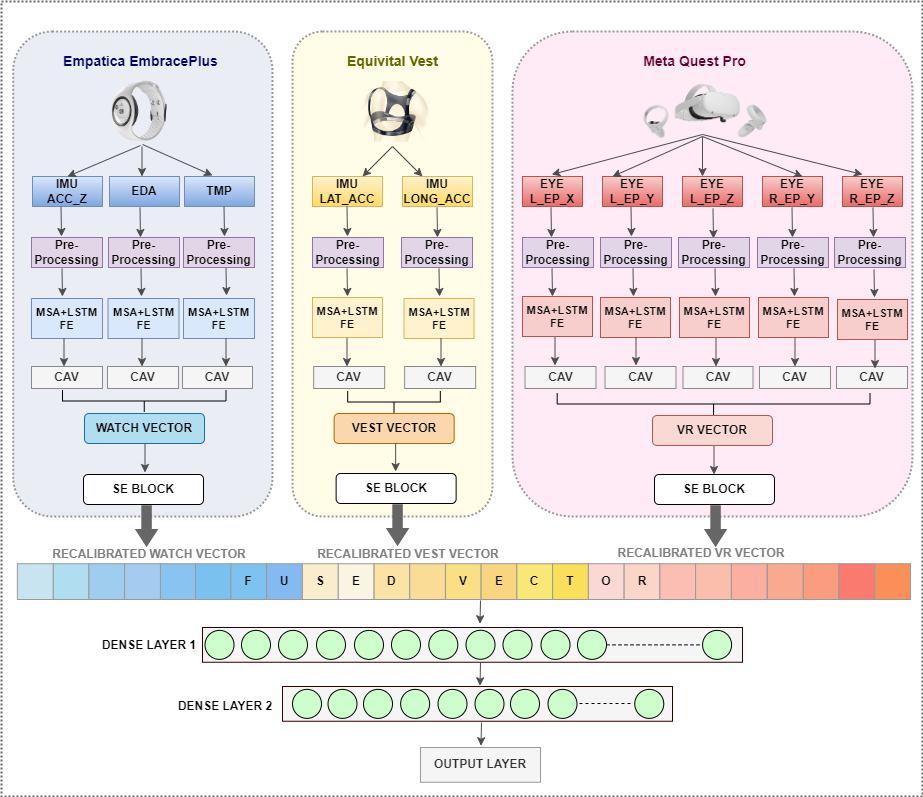}
	\caption{Proposed multi-domain leveraged multimodal deep learning architecture for emotion (valance, arousal) classification using multi-scale attention, LSTM models and squeeze and excitation block. Only best performing modalities from each domain are outlined.}
	\label{arc}
	\vspace{-0.1in}
\end{figure*}
In contrast, IMU data involves motion-related signals captured by accelerometers, which measure acceleration along three axes x, y, and z on the watch, providing insights into movement dynamics. In addition to the Empatica EmbracePlus data, the Equivital Vest offers complementary accelerometer measurements, including the lateral accelerometer, which measures side-to-side movement, the longitudinal accelerometer, which captures forward and backward motion, and the vertical accelerometer, which tracks up-and-down movements. Together, these physiological and motion-related signals provide a comprehensive understanding of the user’s physical and emotional states.

The VR headset predominantly tracks eye movement, with left and right eye positions (LEyeposition, REyeposition) measured in three-dimensional space (x, y, z coordinates). These eye position signals provide precise spatial coordinates at specific time points, enabling us to capture gaze dynamics and spatial orientation during each experimental task.

For peripheral domain signal preprocessing, we applied tailored filtering techniques to ensure high-quality data. Accelerometer signals were filtered using a 0.5-20 Hz band-pass filter to eliminate low-frequency drift and high-frequency noise, preserving relevant motion information. Regarding EDA preprocessing, upsampling from 4 to 64 Hz was performed to make both signals at the equal sampling frequency \cite{chandra2021comparative}. A Butterworth filter with a cutoff frequency of 0.5 Hz is often used. Apply a bandpass filter between 0.5 Hz and 4 Hz to remove noise for BVP signal processing. Use a moving average to reduce the noise from the temperature readings.

For trunk domain data, ECG signals, a 0.5-45 Hz band-pass filter was used to retain key waveform features (P, QRS, T waves) while removing artifacts and noise. Same as the watch, accelerometer signals are filtered using a 0.5-20 Hz band-pass filter. However, GSR data were unreliable for five participants; removing them would cause data insufficiency. Thus, GSR was excluded to maintain consistency across participants.

For each physiological, IMU, and eye-tracking data input, we first applied z-score normalization to standardize the data, ensuring that all signals were on a comparable scale. Following this, we selected the last 40 seconds of data from both the peripheral and trunk domains. This selection was based on an analysis of the signals, which indicated that the final segment of the video stimuli most consistently elicited emotional responses. The extraction of this uniform 40-second segment was necessary to maintain equal input lengths across modalities, a critical requirement for the proposed deep learning architecture.

After segmentation, we applied a sliding window technique with a 2-second window and a 50\% overlap to the physiological and IMU signals. This windowing strategy effectively partitioned the input data stream into multiple overlapping windows, ensuring fine-grained temporal resolution while preserving temporal context. Each segmented window was then represented as a 2D array with dimensions $T \times F$, where $T$ is the number of time windows (time stamps), and $F$ represents the size of the data (features) in one window, determined by the sampling frequency and window duration. Specifically, for signals from the Empatica Embrace Plus, $F$ is set to 128, based on its sampling rate of 64 Hz, while for the Equivital vest, $F$ is 512, reflecting its sampling rate of 256 Hz.

For the Meta Quest Pro VR headset data,  we extracted the final 2000 coordinates from each axis of both the left and right eye positions (x, y, z for each eye). The data was then segmented into overlapping windows using a sliding window approach, with 50\% overlap, dividing the 2000 coordinates into windows of $F$=200 coordinates each. This segmentation process provided structured 2D inputs for the proposed architecture, similar to the representation used for the physiological and IMU data, ensuring compatibility across all input modalities.

\section{System Architecture}

The high-level structure of our proposed emotion recognition system, \textbf{EMO-MSASE}: Emotion Detection using \underline{M}ulti-\underline{S}cale \underline{A}ttention and \underline{S}queeze-and-\underline{E}xcitation
 is depicted in Fig. \ref{arc}, highlighting only the best-performing signals from each device, as determined through experimental evaluation. The system is composed of three core components: data preprocessing, feature extraction, and emotion recognition. Each of these components plays a crucial role in the overall system, and their detailed descriptions are provided in the subsequent sections. \textbf{EMO-MSASE} enhances the framework presented in \cite{yang2021behavioral}, which used a simple attention mechanism. In contrast, we introduce multi-scale attention (MSA), which allows the model to capture patterns across multiple temporal scales, thereby improving its ability to handle varying signal dynamics. Additionally, the integration of Squeeze-and-Excitation (SE) blocks helps recalibrate features by focusing on the most important features for each domain before constructing the fused feature vector from three domains, further enhancing the model’s performance.


\subsection{Multi-Scale Attention LSTM Feature Extraction }
\vspace{-0.1 cm}
In this stage, we employed an LSTM architecture augmented with a multi-scale attention mechanism to act as a high-level feature extractor. The LSTM model \cite{graves2012long}, a specialized variant of recurrent neural networks (RNNs), is adept at capturing long-term dependencies within sequential data. The core elements of the LSTM structure include the cell state and gating mechanisms; the cell state functions as a “memory” that carries information across time steps, while the gating structures control which information is retained or discarded. Specifically, the LSTM comprises three gate types: the forget gate, input gate, and output gate. The forget gate selectively removes irrelevant information from prior cell states, the input gate updates the cell state with new data from the current input, and the output gate produces the final output based on the updated cell state. Utilizing the LSTM enables us to capture temporal relationships within each input modality, yielding a detailed representation.

Additionally, recognizing that not all sub-samples contribute equally to the final recognition task (for example, not all frames in a video are equally informative), we incorporated an attention mechanism \cite{bahdanau2014neural}. This approach assigns relative importance weights to different sub-samples, allowing for the fusion of these weighted sub-samples into a final informative feature vector. The attention mechanism thereby enhances our ability to capture both temporal and spatial dependencies by differentially weighting each sub-sample, while the fused feature vector, being more compact, also reduces training time requirements \cite{zhang2019inferring}.

\begin{figure}
	\centering
	\includegraphics[width=.9\linewidth]{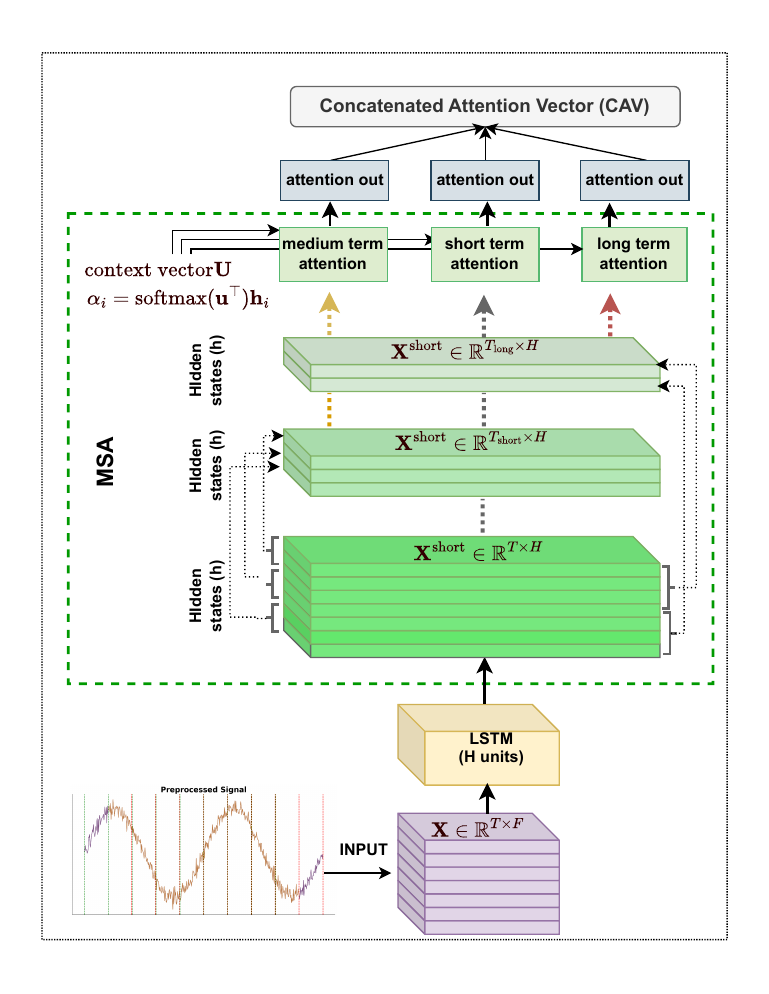}
	\caption{Proposed multi-attention (MSA) based LSTM feature extractor for a modality.}
	\label{f04}
	\vspace{-0.1in}
\end{figure}


The proposed architecture is structured around a multi-domain, multi-modality framework, incorporating three distinct domains: the trunk, peripheral (wrist), and head, represented by the Equivital Vest, Empatica Embrace Plus Watch, and Meta Quest Pro VR headset, respectively. We denote these domains as \( \mathcal{D}_V \) (Equivital Vest), \( \mathcal{D}_W \) (Empatica Embrace Plus), and \( \mathcal{D}_H \) (Meta Quest Pro). Each domain-specific device has multiple modalities to capture diverse physiological and motion-related signals across these domains. For the watch domain, the modalities are denoted as \( \mathbf{X}_W^i \), where \( i \) indexes the different modalities, for example, accelerometer data in the x, y, and z directions. Similarly, the modalities for the vest domain are represented as \( \mathbf{X}_V^j \) and for the VR headset as \( \mathbf{X}_H^k \), with \( j \) and \( k \) indexing the respective modalities within each domain.

Each modality follows a consistent feature extraction process using an LSTM + multi-scale attention (MSA) architecture as shown in Fig. 4. Let \( X \in \mathbb{R}^{T \times F} \) represent the input signal for a given modality, where \( T \) is the number of timesteps and \( F \) is the number of features per timestep. The LSTM processes this sequence and outputs a sequence of hidden states \( H \in \mathbb{R}^{T \times H} \), where \( H \) is the size of the hidden state. Mathematically, the output of the LSTM can be represented as:

\[
H = \text{LSTM}(X) = \begin{bmatrix} h_1 \\ \vdots \\ h_T \end{bmatrix}, \quad h_i \in \mathbb{R}^H
\]

Here, \( h \) denotes the hidden state vector at each timestep, capturing features extracted by the LSTM.

The multi-scale attention (MSA) mechanism is then applied to the LSTM output, operating at three different temporal scales: short-term, medium-term, and long-term. The attention weights for each scale are generated by comparing the hidden states to a learnable context vector, \( \mathbf{u} \in \mathbb{R}^H \), which is initialized randomly and trained to prioritize the most informative timesteps for each modality. The attention score \( \alpha_i \) for each hidden state \( h_i \) is computed as:
\[
\alpha_i = \frac{\exp(\mathbf{u}^\top \cdot h_i)}{\sum_{j=1}^{T} \exp(\mathbf{u}^\top \cdot h_j)}
\]

The learnable context vector \( \mathbf{u} \) helps highlight important information by guiding the focus on particular timesteps based on the context of the task.

For the short-term attention, attention weights \( \alpha_i \) are computed over all \( T \) timesteps, producing a weighted sum of the hidden states:
\[
\mathbf{v}_{\text{short}} = \sum_{i=1}^{T} \alpha_i \mathbf{h}_i, \quad \mathbf{v}_{\text{short}} \in \mathbb{R}^{H}.
\]
For the medium-term attention, adjacent timesteps are merged, reducing the sequence length to \( T_{\text{medium}} = \frac{T}{2} \), and attention is applied over the merged sequence with weights \( \beta_i \):
\[
\mathbf{v}_{\text{medium}} = \sum_{i=1}^{T_{\text{medium}}} \beta_i \mathbf{h}_{\text{merged}, i}, \quad \mathbf{v}_{\text{medium}} \in \mathbb{R}^{H}.
\]
Similarly, for long-term attention, the sequence is further reduced to \( T_{\text{long}} = \frac{T}{3} \), and attention is applied over this coarser representation with weights \( \gamma_i \):
\[
\mathbf{v}_{\text{long}} = \sum_{i=1}^{T_{\text{long}}} \gamma_i \mathbf{h}_{\text{merged}, i}, \quad \mathbf{v}_{\text{long}} \in \mathbb{R}^{H}.
\]

The outputs from the short-term, medium-term, and long-term attention mechanisms are concatenated to form the final feature vector, called the Concatenated Attention Vector (CAV) for the modality:
\[
\text{CAV} = \mathbf{v}_{\text{modality}} = [\mathbf{v}_{\text{short}}, \mathbf{v}_{\text{medium}}, \mathbf{v}_{\text{long}}], \quad \mathbf{v}_{\text{modality}} \in \mathbb{R}^{3H}.
\]

Within each domain, all modality-specific feature vectors are concatenated to form a domain-specific feature vector. For example, in the watch domain:
\[
\mathbf{v}_W = [\mathbf{v}_W^1, \mathbf{v}_W^2, \dots, \mathbf{v}_W^{M_W}], \quad \mathbf{v}_W \in \mathbb{R}^{M_W \times 3H},
\]
where \( M_W \) is the number of modalities in the watch domain. The same process is followed for the vest domain (\( \mathbf{v}_V \)) and the VR headset domain (\( \mathbf{v}_H \)).

After concatenating the modality-specific feature vectors for each domain, a Squeeze-and-Excitation (SE) block \cite{hu2018squeeze} is applied to each concatenated domain-specific feature vector (\( \mathbf{v}_{\text{device}} \)). First, a global average pooling operation is applied to the concatenated domain-specific feature vector:
\[
\mathbf{z}_{\text{device}} = \text{GlobalAvgPool}(\mathbf{v}_{\text{device}}), \quad \mathbf{z}_{\text{device}} \in \mathbb{R}^{3H}.
\]
\( \mathbf{z}_{\text{device}} \) is the result of this global average pooling applied to the concatenated feature vector \( \mathbf{v}_{\text{device}} \) for a specific device (watch, vest, or VR headset).

Next, two fully connected layers with ReLU and sigmoid activations generate recalibration weights for each device:
\[
\mathbf{s}_{\text{device}} = \sigma(\mathbf{W}_2 \cdot \text{ReLU}(\mathbf{W}_1 \cdot \mathbf{z}_{\text{device}})), \quad \mathbf{s}_{\text{device}} \in \mathbb{R}^{3H}.
\]

\( \mathbf{s}_{\text{device}} \) represents the recalibration weights generated for a specific device after applying two fully connected layers to \( \mathbf{z}_{\text{device}} \). The recalibration weights are then applied to the device-specific feature vector:
\[
\mathbf{v}_{\text{device}}^{\text{SE}} = \mathbf{s}_{\text{device}} \cdot \mathbf{v}_{\text{device}}, \quad \mathbf{v}_{\text{device}}^{\text{SE}} \in \mathbb{R}^{M_W \times 3H}.
\]

The recalibrated domain-specific feature vectors are then concatenated to form a global feature vector:
\[
\mathbf{v}_{\text{global}} = [\mathbf{v}_W^{\text{SE}}, \mathbf{v}_V^{\text{SE}}, \mathbf{v}_H^{\text{SE}}], \quad \mathbf{v}_{\text{global}} \in \mathbb{R}^{(M_W + M_V + M_H) \times 3H}.
\]

Finally, the global feature vector is passed through a fully connected layer for classification, with softmax activation for multi-class classification:
\[
\hat{\mathbf{y}} = \text{softmax}(\mathbf{W}_{\text{output}} \cdot \mathbf{v}_{\text{global}} + \mathbf{b}_{\text{output}}), \quad \hat{\mathbf{y}} \in \mathbb{R}^C.
\]
Here, \( \hat{\mathbf{y}} \) represents the predicted class probabilities for \( C \) emotion classes.

\subsection{Deep Learning Setup For the Proposed Architecture}
For feature extraction, each device utilizes a dedicated Attention-LSTM network. The LSTM network comprises two layers, each with 128 hidden units. The model is trained using binary cross-entropy loss with the AdamW optimizer for effective weight updates, a learning rate of 0.001, and a batch size of 16. Training is conducted over 50 epochs, with early stopping based on validation performance to prevent overfitting. All data analytics and deep learning model building and testing were conducted on a 13th-generation i7 workstation equipped with an RTX 4080 GPU and 32GB of RAM.

\section{results}
\begin{figure*}[ht]
  \centering
  \begin{minipage}{0.48\textwidth}
    \centering
    \includegraphics[width=\textwidth]{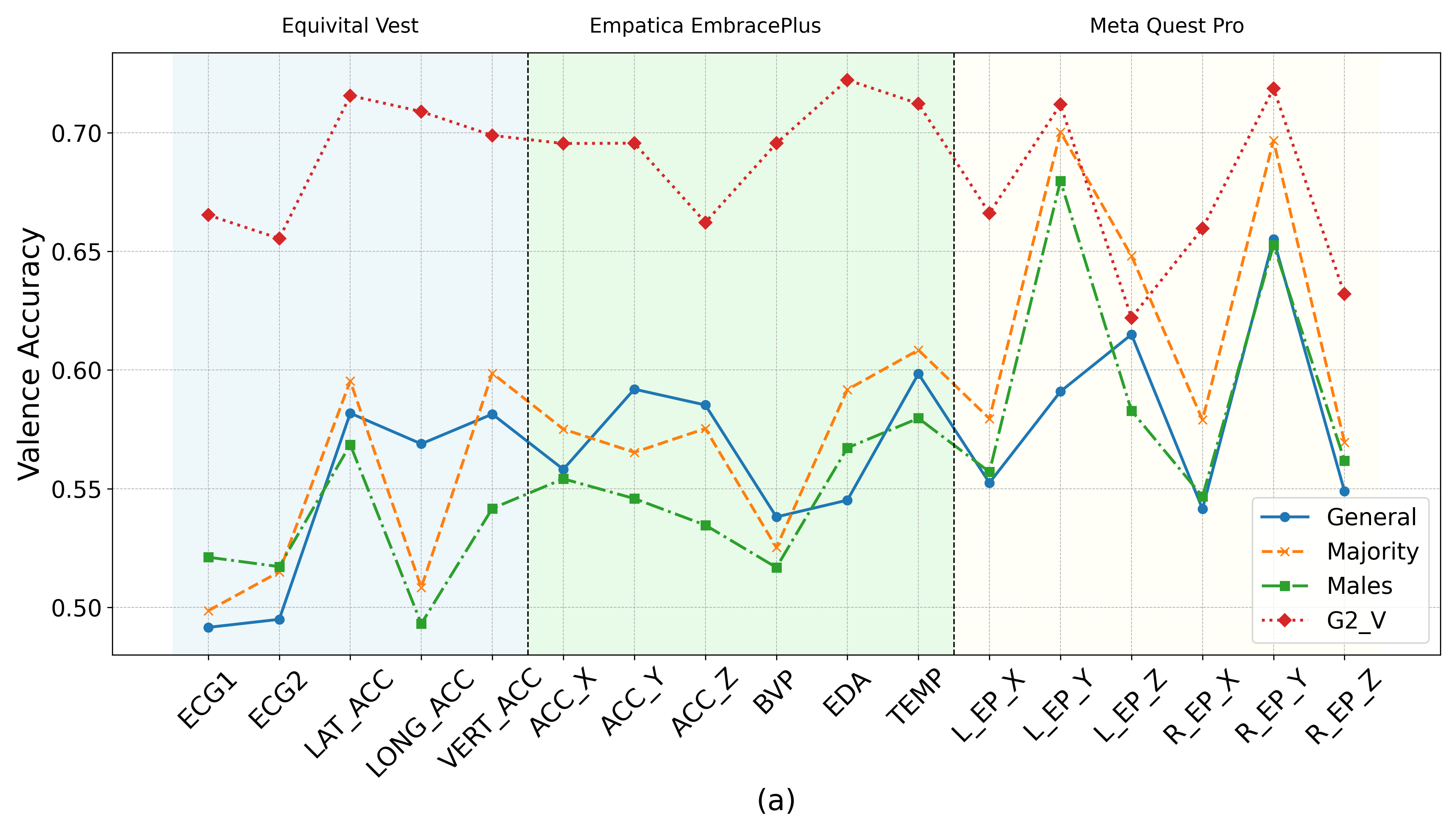}
    \label{fig:valence}
  \end{minipage}
  \hfill
  \begin{minipage}{0.48\textwidth}
    \centering
    \includegraphics[width=\textwidth]{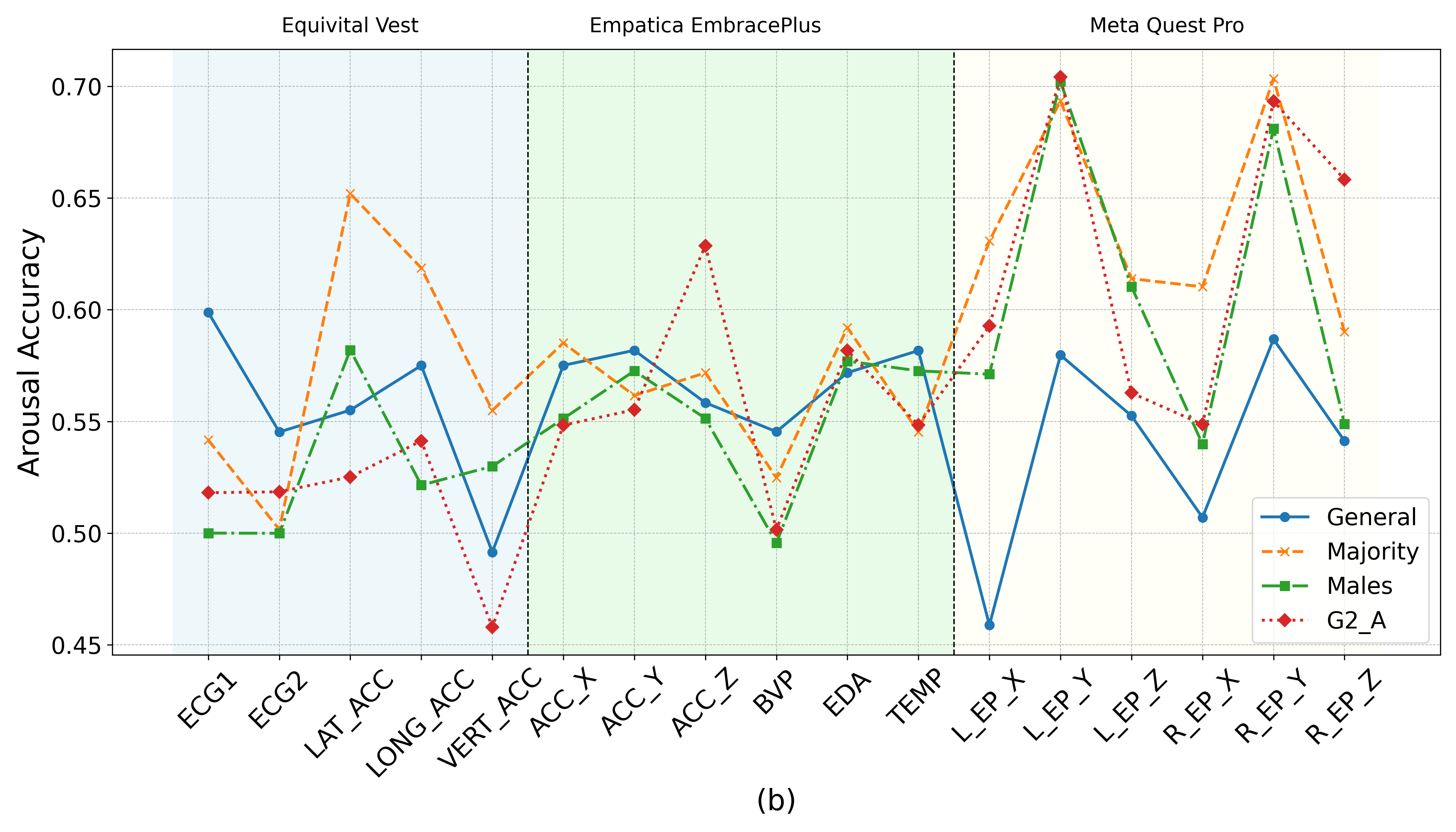}
    \label{fig:arousal}
  \end{minipage}
  \caption{Comparison of unimodal valence and arousal accuracy across four different cases (general, majority, males only and G2) for three domains. Blue color indicates the trunk domain, green color indicates the peripheral domain and red color indicates head domain, (a) shows the valence comparison, and (b) shows the arousal comparison.}
  \label{fig:comparison}
\end{figure*}

In this section, we present the performance of \textbf{EMO-MSASE}. Our study leverages data from three distinct domains, head (Meta Quest Pro VR headset), trunk (Equivital Vest), and peripheral (Empatica Embrace Plus watch) to advance emotion recognition. These domains contribute multiple modalities, including IMU data, ECG, temperature, EDA, BVP, and eye movements as in Table \ref{ref:domain info}, which are utilized to assess valence and arousal. While these devices offer a wide array of signals, we selected only the most relevant modalities for our analysis, focused on their significance in emotion detection. We collected a total of 299 samples for each modality from 23 participants. Each participant watched thirteen videos, completing the SAM Manikin Questionnaire after each video to rate their emotional responses on a scale from 1 to 7 for both valence and arousal. As indicated in Table \ref{ref:video selection}, we classified ratings  $ < 4 $ as low valence (LV) and low arousal (LA), and ratings $ > 4 $ as high valence (HV) and high arousal (HA), forming the \textit{first ground truth} (G1) for valence (G1\_V) and arousal (G1\_A) respectively. The \textit{second ground truth} (G2) of valence (G2\_V) and arousal (G2\_A) levels is derived from external ratings (three domain experts, two psychologists and one computer scientist) based on established criteria from previous research \cite{guimard2022pem360,voigt2020comparing, li2017public}. In constructing the deep learning model, four distinct cases of label assignment were systematically applied to the collected dataset: (1) the actual response of each participant for each video was used as the label (general), (2) for each video, a final label was determined by the majority response across all 23 participants (majority) as in Table \ref{ref:video selection}, (3) only the responses from male participants were used (males only), and (4) the label values provided by the selected reference papers (G2\_V and G2\_A) for each video to assign labels for each video. Cases (1), (2), and (3) are subclasses of the general labels (G1\_V and G1\_A).

The efficacy of our proposed deep learning architecture for valence and arousal detection was rigorously evaluated through two distinct cross-validation techniques: (1) Group K-Fold cross-validation and (2) Leave-One-Subject-Out (LOSO) cross-validation. In Group K-Fold cross-validation, participants were divided into two distinct groups for training and testing, ensuring that data from the same participant was not used in both the training and testing sets to prevent data leakage. This approach maintains subject independence during model evaluation. In LOSO cross-validation, the model was trained on data from all but one participant, with the left-out participant's data used for testing. This process was repeated for each participant, allowing for a robust assessment of the model’s generalizability across individuals. For Analysis \ref{analysis_1}, \ref{analysis_2}, \ref{analysis_3}, we employed Group
K-Fold cross-validation with 5 folds. The mean accuracy across the 5 folds was computed to provide a robust and
comprehensive evaluation of the model’s performance. To further enhance the evaluation, Analysis \ref{analysis_3} was additionally assessed using (LOSO) cross-validation.

\subsection{Uni-modality Performance Analysis}
\label{analysis_1}
We began by independently evaluating the performance of each signal modality within each domain, as shown in Table \ref{ref:domain info}, for the detection of valence and arousal. Fig \ref{fig:comparison} illustrates the results of individual signal performance for both valence (a) and arousal (b) cross four distinct cases. In the context of valence detection, the model demonstrates robust performance across almost all signals from the three domains when using G\_2 labels, consistently achieving over 70\% accuracy. Notably, LAT\_ACC from the trunk, EDA and TEMP from the peripheral domain, and L\_EP\_Y and R\_EP\_Y from the head domain perform particularly well under G\_2 labeling. However, when using G\_1 labels, the signals from the trunk and peripheral domains exhibit lower accuracy across all three scenarios (general, majority, and males). Upon closer examination, the accuracy for trunk and peripheral signals is slightly improved under the majority labeling case compared to the other two. In contrast, the head domain exhibits significantly higher performance when majority labeling is employed. Notably, in this scenario, head domain signals outperform those from the trunk and peripheral domains. Specifically, L\_EP\_Y and R\_EP\_Y independently achieve close to 70\% accuracy for valence detection, indicating strong predictive power for these signals in alignment with the majority labeling.

Unlike valence, arousal detection demonstrates superior performance with the majority labeling across most signals from all three domains. A similar trend is observed for both valence and arousal in the head domain, where L\_EP\_Y and R\_EP\_Y independently achieve nearly 70\% accuracy. In contrast, G2\_V labeling yields comparatively lower performance for signals in the trunk and peripheral domains, though the head domain performance remains consistent with that of the majority labeling. LAT\_ACC emerges as the most predictive signal in the trunk domain, while EDA stands out as the best-performing signal in the peripheral domain.

In conclusion, majority labeling consistently performs well for both valence and arousal detection when applied to our SAM ratings.
\begin{table*}[]
\tiny
\caption{Performance of different combinations for Valence and Arousal detection using the Peripheral (Empatica EmbracePlus) signals across the following cases: General, Majority, Males, G2\_V, and G2\_A.}
\resizebox{\textwidth}{!}{%
\begin{tabular}{lcccccccc}
\hline
\multirow{2}{*}{\textbf{Peripheral signal Combinations}} & \multicolumn{4}{c}{\textbf{Valence}} & \multicolumn{4}{c}{\textbf{Arousal}} \\ \cline{2-9} 
 & \textbf{General} & \textbf{Majority} & \textbf{Males} & \textbf{G2\_V} & \textbf{General} & \textbf{Majority} & \textbf{Males} & \textbf{G2\_A} \\ \hline
ACC\_Z, TEMP & 0.558531 & 0.615173 & 0.542831 & \textbf{0.735593} & 0.561695 & 0.588763 & 0.555597 & \textbf{0.618701} \\ \hline
ACC\_X, ACC\_Y, EDA & 0.555028 & 0.541808 & 0.551434 & 0.692034 & 0.524802 & 0.548305 & 0.598150 & 0.528418 \\ \hline
ACC\_X, ACC\_Z, TEMP & 0.551921 & 0.578435 & 0.555967 & 0.692316 & 0.565198 & 0.564365 & 0.538483 & 0.568362 \\ \hline
ACC\_Z, EDA, TEMP & 0.568475 & \textbf{0.618365} & 0.521739 & 0.735819 & 0.598701 & \textbf{0.611476} & 0.581499 & 0.605254 \\ \hline
\end{tabular}%
}
\label{watch}
\end{table*}

\begin{table*}[]
\tiny
\caption{Performance of different combinations for Valence and Arousal detection using the Trunk (Equivital Vest) signals across the following cases: General, Majority, Males, G2\_V, and G2\_A.}
\resizebox{\textwidth}{!}{%
\begin{tabular}{lcccccccc}
\hline
\multirow{2}{*}{\textbf{Trunk signal Combinations}} & \multicolumn{4}{c}{\textbf{Valence}} & \multicolumn{4}{c}{\textbf{Arousal}} \\ \cline{2-9} 
 & \textbf{General} & \textbf{Majority} & \textbf{Males} & \textbf{G2\_V} & \textbf{General} & \textbf{Majority} & \textbf{Males} & \textbf{G2\_A} \\ \hline
LAT\_ACC, LONG\_ACC & \textbf{0.581808} & 0.531864 & 0.547179 & \textbf{0.715367} & 0.534972 & \textbf{0.608701} & 0.555967 & \textbf{0.577243} \\ \hline
ECG1, LAT\_ACC, LONG\_ACC & 0.545367 & 0.532203 & 0.530250 & 0.692203 & 0.548418 & 0.565311 & 0.611286 & 0.517576 \\ \hline
ECG2, LAT\_ACC, LONG\_ACC & 0.531695 & 0.501864 & 0.555504 & 0.695537 & 0.531864 & 0.522090 & 0.547271 & 0.542831 \\ \hline
LAT\_ACC, LONG\_ACC, VERT\_ACC & 0.541695 & 0.589765 & 0.568494 & 0.668870 & 0.505141 & 0.558431 & 0.572800 & 0.530342 \\ \hline
\end{tabular}%
}
\label{vest}
\end{table*}

\begin{table*}[]
\caption{Performance of different combinations for Valence and Arousal detection using the Head (Meta Quest Pro VR Head Set) signals across the following cases: General, Majority, Males, G2\_V, and G2\_A.}
\resizebox{\textwidth}{!}{%
\begin{tabular}{lcccccccc}
\hline
\multirow{2}{*}{\textbf{VR headset modality Combination}} & \multicolumn{4}{c}{\textbf{Valence}} & \multicolumn{4}{c}{\textbf{Arousal}} \\ \cline{2-9} 
 & \textbf{General} & \textbf{Majority} & \textbf{Males} & \textbf{G2\_V} & \textbf{General} & \textbf{Majority} & \textbf{Males} & \textbf{G2\_A} \\ \hline
L\_EP\_X, R\_EP\_Y & 0.563354 & 0.659614 & 0.601836 & 0.721625 & 0.518995 & 0.694155 & 0.645507 & 0.632729 \\ \hline
L\_EP\_Y, L\_EP\_Z & 0.673290 & 0.687317 & 0.646087 & 0.714728 & 0.594565 & 0.697487 & 0.689758 & 0.697896 \\ \hline
L\_EP\_Y, R\_EP\_X & 0.556458 & 0.646172 & 0.615169 & 0.742139 & 0.522385 & 0.683869 & 0.654396 & 0.666686 \\ \hline
L\_EP\_Y, R\_EP\_Z & 0.566920 & 0.663179 & 0.645700 & 0.731736 & 0.539509 & 0.676856 & 0.614686 & 0.677089 \\ \hline
L\_EP\_X, L\_EP\_y, L\_EP\_Z & 0.614787 & 0.676973 & 0.645797 & 0.721566 & 0.525833 & 0.704383 & 0.685507 & 0.714904 \\ \hline
L\_EP\_X, L\_EP\_Y, R\_EP\_Y & 0.611338 & 0.663063 & 0.619517 & 0.742314 & 0.553361 & 0.676973 & 0.654300 & 0.667037 \\ \hline
L\_EP\_Y, L\_EP\_Z, R\_EP\_Y & 0.614728 & 0.673641 & 0.668019 & 0.735301 & 0.556400 & 0.690649 & 0.667633 & 0.694506 \\ \hline
L\_EP\_Y, L\_EP\_Z, R\_EP\_Z & 0.611572 & 0.670076 & 0.628213 & 0.731794 & 0.563238 & 0.680245 & 0.658744 & 0.694448 \\ \hline
L\_EP\_Y, R\_EP\_Y, R\_EP\_Z & 0.608007 & 0.666628 & 0.628213 & \textbf{0.752367} & 0.563530 & 0.690824 & 0.663575 & 0.694272 \\ \hline
R\_EP\_X, R\_EP\_Y, R\_EP\_Z & 0.587317 & 0.625132 & 0.632560 & 0.741905 & 0.511864 & 0.656341 & 0.641159 & 0.683811 \\ \hline
L\_EP\_X, L\_EP\_Y, L\_EP\_Z, R\_EP\_Y & 0.639041 & 0.697545 & 0.650338 & 0.731911 & 0.580713 & 0.725015 & 0.671884 & 0.694331 \\ \hline
L\_EP\_X, L\_EP\_Y, R\_EP\_Y, R\_EP\_Z & 0.604383 & 0.680245 & 0.637198 & 0.745587 & 0.539626 & 0.639217 & 0.632367 & 0.708182 \\ \hline
L\_EP\_X, L\_EP\_Y, L\_EP\_Z, R\_EP\_Y, R\_EP\_Z & 0.635535 & \textbf{0.710570} & 0.659324 & 0.735126 & 0.563530 & \textbf{0.707773} & 0.658744 & \textbf{0.732262} \\ \hline
L\_EP\_Y, L\_EP\_Z, R\_EP\_X, R\_EP\_Y, R\_EP\_Z & 0.587317 & 0.669959 & 0.610628 & 0.735243 & 0.560023 & 0.683694 & 0.641063 & 0.721917 \\ \hline
\end{tabular}%
}
\label{vr}
\end{table*}

\subsection{Single Domain Performance Analysis}
\label{analysis_2}

In this experiment, we conducted an independent evaluation of each domain's performance using their respective modalities for valence and arousal detection. For the peripheral domain, we initially considered four key modalities: IMU data (accelerometer readings along the X, Y, and Z axes), BVP, EDA, and TEMP. These six raw signals across four modalities yielded 64 possible combinations. Table \ref{watch} showcases the top four combinations for both valence and arousal detection. For the trunk domain, we analyzed two key modalities: electrocardiogram (ECG) and IMU data. The ECG data was captured from two leads, referred to as ECG1 and ECG2, while the IMU data included lateral acceleration (LAT\_ACC), longitudinal acceleration (LONG\_ACC), and vertical acceleration (VERT\_ACC). The top five performing combinations for valence and arousal detection are summarized in Table \ref{vest}. For the eye-tracking data, under the eye modality, we utilized  six features: LEyeposition\_x, LEyeposition\_y, LEyeposition\_z, REyeposition\_x, REyeposition\_y, and REyeposition\_z. The results of the top-performing combinations are presented in Table \ref{vr}.

As shown in Tables \ref{watch} and \ref{vest}, higher accuracy values for both valence and arousal were consistently observed with G2\_V and G2\_A labeling, except in the trunk domain. For arousal detection, the trunk domain achieved the highest accuracy under majority labeling. In contrast, when using G1\_V and G1\_A labels, valence accuracy was moderate across all three cases (general, majority, and males) in both the peripheral and trunk domains. Overall, majority labeling yielded the highest accuracy for both valence and arousal in the trunk and peripheral domains when using G1\_V and G1\_A labels from the collected data. In analyzing the top-performing signals combinations for the watch, the combination of ACC\_Z, EDA, and TEMP consistently demonstrated superior performance across all cases. Similarly, for the vest, the combination of LAT\_ACC and LONG\_ACC emerged as the most effective, yielding robust results across all conditions.

For the eye-tracking data from the VR headset, we achieved significantly higher valence and arousal accuracy compared to other domains across all four cases as in Table \ref{vr}. In the general case, valence accuracy for the other domains was 55\% ± 3, while the head data yielded a notably higher accuracy of 60\% ± 3. Under majority labeling, valence accuracy remained at 57\% ± 4, whereas the head data improved substantially to 66\% ± 3. For male participants, valance accuracy was 54\% ± 3, compared to a significantly higher 64\% ± 3 for head data. Arousal accuracy followed a similar trend, with 55\% ± 4 in the general case and 57\% ± 3 for the head domain. In the majority case, the accuracy was 57\% ± 4 for the peripheral and 56\% ± 4 for the trunk, while the eye-tracking data demonstrated a significant improvement, achieving 68\% ± 3. For male participants, arousal accuracy was 56\% ± 5 for both the peripheral and trunk, but notably higher for the eye-tracking data at 65\% ± 3. The optimal signal combination accross all cases for the head domain was found to consist of five key eye signals: LEyeposition\_x, LEyeposition\_y, LEyeposition\_z, REyeposition\_y, and REyeposition\_z.

Upon analyzing the final outcomes from all three domains, we observed that, for both the peripheral and trunk domains, the majority case performed slightly better than the general and males-only cases. However, for the VR data, the majority case demonstrated significantly superior performance. Based on these findings, we selected the best-performing modality combinations from each domain and utilized them to design the multi-domain architecture, focusing exclusively on the majority case. The results of this multi-domain architecture are presented in the following subsection.

\subsection{Multi-Domain Fusion Performance Analysis}
 The results presented in Table \ref{ref:multi-new} clearly demonstrate the superiority of multi-domain fusion over single-domain approaches in both valence and arousal detection.
\label{analysis_3}
\begin{table}[H]
\tiny
\caption{Performance of modality combinations from Watch, Vest, and VR for Valence (G1\_V, G2\_V) and Arousal (G1\_A, G2\_A) detection using \textbf{EMO-MSASE} in Fig. \ref{arc}}
\resizebox{\columnwidth}{!}{%
\begin{tabular}{lcccc}
\hline
\textbf{Combination} & \textbf{G1\_V} & \textbf{G2\_V} & \textbf{G1\_A} & \textbf{G2\_A} \\ \hline
Peripheral & 61.83 & 73.58 & 58.87 & 61.87 \\ \hline
Trunk & 53.18 & 71.53 & 60.87 & 57.72 \\ \hline
Head & 71.05 & 73.51 & \textbf{70.77} & \textbf{73.22} \\ \hline
Peripheral + Trunk & 58.13 & 82.76 & 60.33 & 59.76 \\ \hline
Peripheral + Head & 72.99 & 81.96 & 62.86 & 60.54 \\ \hline
Trunk + Head & 77.89 & \textbf{87.14} & 63.51 & 61.23 \\ \hline
Peripheral + Trunk + Head & \textbf{80.57} & 86.77 & 65.15 & 62.53 \\ \hline
\end{tabular}%
}
\label{ref:multi-new}
\end{table}

 While the head domain consistently outperforms both the peripheral and trunk domains individually, achieving accuracies as high as 71.05\% for G1\_V and 73.22\% for G2\_A, the fusion of multiple domains significantly enhances performance. Notably, the combination of trunk and head domains delivers the highest accuracy for G2\_V, reaching 87.14\%, while the full fusion of all three domains (peripheral, trunk, and head) results in the best overall performance for valence detection in G1\_V, with 80.57\%. For arousal detection, the head domain alone achieves the best individual performance, but fusion strategies, particularly those combining head and trunk signals, yield incremental improvements across most cases. These results underscore the effectiveness of integrating signals from different domains, particularly the critical role of head domain signals in achieving optimal accuracy, further validating the strength of the proposed \textbf{EMO-MSASE} architecture.

\subsection{Ablation Study}
Table \ref{ablation com} presents the results of the ablation experiments conducted on the proposed multimodal fusion framework. In these experiments, LSTMSA denotes the use of self-attention with LSTM for feature extraction, LSTMMSA represents the application of a multi-scale attention mechanism with LSTM in the fusion layer, and proposed architecture, \textbf{EMO-MSASE} indicates the integration of both multi-scale attention and a Squeeze-and-Excitation (SE) block for each domain prior to domain fusion.

The results provide clear evidence of the benefits introduced by the \textbf{EMO-MSASE} architecture compared to the LSTMSA and LSTMMSA models. Across all domain combinations and both valence and arousal detection, \textbf{EMO-MSASE} consistently outperforms the other models. For valence detection, the fusion of trunk and head signals achieves the highest accuracy of 87.14\% for G2\_V, significantly surpassing the best results from LSTMMSA (84.28\%) and LSTMSA (80.56\%). Similarly, in arousal detection, \textbf{EMO-MSASE} delivers superior performance, with 65.15\% accuracy for G1\_A and 63.53\% for G2\_A when combining all domains, again outperforming LSTMMSA and LSTMSA. The addition of the multi-scale attention mechanism, along with the Squeeze-and-Excitation (SE) blocks, enhances the model’s capacity to capture complex patterns across domains, leading to the observed improvements in accuracy. Notably, the combination of head and peripheral domains in \textbf{EMO-MSASE} also shows strong performance, particularly in valence detection, highlighting the importance of multi-modal fusion in this task. These results underline the efficacy of integrating domain-specific SE blocks and multi-scale attention, positioning \textbf{EMO-MSASE} as the most effective model in the presented framework.


\begin{table*}
\tiny
\caption{Ablation study results for valence (G1\_V, G2\_V) and arousal (G1\_A, G2\_A) detection. The table presents accuracy and recall values for different domain combinations using three models: LSTMSA, LSTMMSA, and our proposed model.}
\resizebox{\textwidth}{!}{%
\begin{tabular}{llcccccccc}
\hline
\multirow{3}{*}{\textbf{Models}} & \multirow{3}{*}{\textbf{Domain Combination}} & \multicolumn{4}{c}{\textbf{Valence}} & \multicolumn{4}{c}{\textbf{Arousal}} \\ \cline{3-10} 
 &  & \multicolumn{2}{c}{\textbf{Accuracy}} & \multicolumn{2}{c}{\textbf{Recall}} & \multicolumn{2}{c}{\textbf{Accuracy}} & \multicolumn{2}{c}{\textbf{Recall}} \\ \cline{3-10} 
 &  & \textbf{G1\_V} & \textbf{G2\_V} & \textbf{G1\_V} & \textbf{G2\_V} & \textbf{G1\_A} & \textbf{G2\_A} & \textbf{G1\_A} & \textbf{G2\_A} \\ \hline
LSTMSA & \begin{tabular}[c]{@{}l@{}}Head\\ Peripheral + Head\\ Trunk + Head\\ Peripheral + Trunk + Head\end{tabular} & \begin{tabular}[c]{@{}c@{}}64.34\\ 68.23\\ 73.25\\ 74.35\end{tabular} & \begin{tabular}[c]{@{}c@{}}69.72\\ 75.65\\ 80.23\\ 80.56\end{tabular} & \begin{tabular}[c]{@{}c@{}}62.65\\ 67.53\\ 73.19\\ 72.65\end{tabular} & \begin{tabular}[c]{@{}c@{}}67.54\\ 74.76\\ 81.25\\ 79.34\end{tabular} & \begin{tabular}[c]{@{}c@{}}68.54\\ 57.43\\ 58.87\\ 60.54\end{tabular} & \begin{tabular}[c]{@{}c@{}}68.54\\ 55.45\\ 56.35\\ 58.87\end{tabular} & \begin{tabular}[c]{@{}c@{}}67.54\\ 58.87\\ 58.87\\ 59.87\end{tabular} & \begin{tabular}[c]{@{}c@{}}68.53\\ 56.73\\ 57.16\\ 57.76\end{tabular} \\ \hline
LSTMMSA & \begin{tabular}[c]{@{}l@{}}Head\\ Peripheral + Head\\ Trunk + Head\\ Peripheral + Trunk + Head\end{tabular} & \begin{tabular}[c]{@{}c@{}}67.21\\ 70.16\\ 75.25\\ 78.58\end{tabular} & \begin{tabular}[c]{@{}c@{}}72.43\\ 78.87\\ 83.34\\ 84.28\end{tabular} & \begin{tabular}[c]{@{}c@{}}66.35\\ 71.26\\ 73.87\\ 76.94\end{tabular} & \begin{tabular}[c]{@{}c@{}}72.43\\ 76.87\\ 84.65\\ 81.95\end{tabular} & \begin{tabular}[c]{@{}c@{}}70.43\\ 59.87\\ 60.54\\ 63.76\end{tabular} & \begin{tabular}[c]{@{}c@{}}70.87\\ 58.87\\ 58.76\\ 61.65\end{tabular} & \begin{tabular}[c]{@{}c@{}}67.86\\ 62.76\\ 62.17\\ 63.54\end{tabular} & \begin{tabular}[c]{@{}c@{}}70.43\\ 59.64\\ 57.75\\ 60.34\end{tabular} \\ \hline
\textbf{EMO-MSASE} & \begin{tabular}[c]{@{}l@{}}Head\\ Peripheral + Head\\ Trunk + Head\\ Peripheral + Trunk + Head\end{tabular} & \begin{tabular}[c]{@{}c@{}}71.05\\ 72.99\\ 77.89\\ 80.57\end{tabular} & \begin{tabular}[c]{@{}c@{}}73.51\\ 81.96\\ 87.14\\ 86.77\end{tabular} & \begin{tabular}[c]{@{}c@{}}68.21\\ 73.41\\ 75.21\\ 78.32\end{tabular} & \begin{tabular}[c]{@{}c@{}}75.71\\ 80.34\\ 85.15\\ 83.23\end{tabular} & \begin{tabular}[c]{@{}c@{}}70.77\\ 62.86\\ 63.51\\ 65.15\end{tabular} & \begin{tabular}[c]{@{}c@{}}73.22\\ 60.54\\ 61.23\\ 63.53\end{tabular} & \begin{tabular}[c]{@{}c@{}}70.45\\ 64.32\\ 63.12\\ 65.19\end{tabular} & \begin{tabular}[c]{@{}c@{}}72.32\\ 61.54\\ 60.32\\ 62.57\end{tabular} \\ \hline
\end{tabular}%
}
\label{ablation com}
\end{table*}

\subsection{LOSO and Different Fusion Technique Results}
As detailed in \ref{model fusion}, we explored various fusion techniques for integrating multi-modal data, with a primary focus on feature-level fusion in our proposed deep learning architecture. To further enhance the analytical robustness, we also employed decision-level fusion methods, utilizing both sum and max strategies. The sum fusion technique aggregates the probabilistic outputs from multiple classifiers, summing the probabilities for each emotion and selecting the label with the highest cumulative probability. In contrast, the max strategy identifies the highest individual probability across classifiers and designates that as the final prediction. In this analysis, we focus exclusively on the scenario involving the combination of all three domains. Additionally, we computed the Leave-One-Subject-Out (LOSO) accuracy score to further assess the model's generalizability. 
\begin{table}[H]
\large
\caption{Performance of three domain combination for Valence (G1\_V, G2\_V) and Arousal (G1\_A, G2\_A) detection using \textbf{EMO-MSASE}. The table includes results for both K-Fold and LOSO evaluation strategies.}
\resizebox{\columnwidth}{!}{%
\begin{tabular}{lcccccccc}
\hline
\textbf{Fusion } & \multicolumn{2}{c}{\textbf{G1\_V}} & \multicolumn{2}{c}{\textbf{G2\_V}} & \multicolumn{2}{c}{\textbf{G1\_A}} & \multicolumn{2}{c}{\textbf{G2\_A}} \\ \cline{2-9} 
 & \textbf{K-Fold} & \textbf{LOSO} & \textbf{K-Fold} & \textbf{LOSO} & \textbf{K-Fold} & \textbf{LOSO} & \textbf{K-Fold} & \textbf{LOSO} \\ \hline
ML & 80.57 & 78.21 & 86.77 & 83.45 & 65.15 & 64.21 & 62.53 & 60.54 \\ \hline
DL-sum & 74.68 & 71.26 & 80.13 & 77.21 & 60.23 & 58.83 & 56.24 & 54.65 \\ \hline
DL-max & 69.25 & 65.76 & 74.21 & 71.76 & 54.76 & 53.87 & 52.76 & 51.45 \\ \hline

\end{tabular}%
}
\label{ref:fusion loso}
\end{table}
The results in Table \ref{ref:fusion loso} highlight the performance loss associated with decision-level fusion (DL) techniques compared to modality-level fusion (ML) for both valence and arousal detection using the \textbf{EMO-MSASE} framework. For valence detection, ML shows superior performance, and switching to DL-sum leads to an accuracy loss of 6.64\% (from 86.77\% to 80.13\%) for G2\_V, while DL-max results in an even larger drop of 12.56\% (from 86.77\% to 74.21\%). Similarly, for arousal detection, ML significantly outperforms both DL methods, with DL-sum and DL-max incurring accuracy losses of 8.91\% and 12.39\%, respectively, when compared to ML for G1\_A. These results demonstrate that while decision-level fusion can still achieve reasonable performance, it introduces notable accuracy degradation compared to the more comprehensive integration provided by modality-level fusion. This reinforces the advantage of ML in achieving more reliable and accurate emotion detection across multiple domains.

\section{Discussion}
In this section, we present key findings derived from our experimental analysis. As shown in Fig. \ref{fig:comparison}, G2\_V labelling yields superior performance for valence detection, whereas G1 with majority labelling proves more effective for arousal detection, outperforming G2\_A. Overall, majority labelling consistently demonstrates robust results for both valence and arousal when considering participant ratings (G1) from the SAM scale. Another critical observation is the superior performance of head domain signals compared to those from the trunk and peripheral domains in both valence and arousal detection. When evaluating single-domain performance, the head domain exhibits markedly better accuracy for both metrics, aligning with the trends observed in unimodal analysis, where majority labelling again outperforms other labelling strategies. For multi-domain performance analysis, we selected the best-performing signal combinations from each domain. The results indicate that combining signals from all three domains yields the highest accuracy for valence detection. However, for arousal, the head domain alone produces the best performance. Furthermore, the ablation study highlights the substantial impact of incorporating multi-scale attention mechanisms and Squeeze-and-Excitation (SE) blocks into the model architecture, significantly enhancing accuracy compared to conventional approaches. Finally, the evaluation of different fusion techniques reveals that modality-level fusion consistently outperforms decision-level fusion strategies within the proposed \textbf{EMO-MSASE} framework, confirming the efficacy of our fusion strategy for multi-modal emotion detection.

\section{Limitations}
This study has certain limitations. First, the dataset size is relatively small, comprising only 23 participants. Given the potential for subjective variability in participant-reported emotions in response to the same video stimuli, the limited number of participants may result in insufficient data for the model to effectively learn and generalize emotion patterns. In future work, we aim to expand the participant pool to mitigate this limitation. Additionally, when collecting data, we grouped the first eight videos into four pairs, having participants watch two videos consecutively and then provide a single emotion rating. For analysis, we applied the same rating to both videos in each pair. However, we recognize that this approach may not fully capture the emotional nuances participants experience, as they might have felt different or mixed emotions while watching each video. In future experiments, we plan to show one video at a time and allow participants to provide distinct ratings for each video, thereby improving the granularity and accuracy of the collected data. Furthermore, while our model incorporates attention mechanisms to enhance performance, there is room for improvement. Future work should focus on developing more rigorous and sophisticated attention mechanisms to further improve the accuracy of emotion detection across multiple modalities.

\section{Conclusion}

This study presented \textbf{EMO-MSASE}, an advanced multimodal deep learning architecture for the classification of valence and arousal. By leveraging multi-scale attention mechanisms and Squeeze-and-Excitation (SE) blocks within a multi-domain framework, our approach demonstrated the capability to capture nuanced temporal features and enhance signal integration across head, trunk, and peripheral domains. Our work encompassed the full workflow, from hardware selection, system setup, and configuration, to participant recruitment, data acquisition, data cleaning, and preprocessing, ensuring a robust and reliable dataset aligned with real-world applications. Through rigorous evaluations, including Group K-Fold and Leave-One-Subject-Out cross-validation, we showed that integrating modalities from these domains significantly improved emotion classification accuracy, particularly when combining head and trunk data for valence and relying on head data for arousal.

Our findings revealed that majority labeling, explained in Table \ref{ref:video selection} consistently enhanced classification accuracy, with valence detection achieving the highest accuracy under expert-based (G2) labeling, while arousal performed optimally with participant ratings (G1). The ablation study further confirmed the impact of multi-scale attention and SE blocks, with \textbf{EMO-MSASE} surpassing traditional LSTM with self-attention (LSTMSA) and LSTM with multi-scale attention (LSTMMSA) approaches in capturing complex, domain-specific patterns.

Moreover, the superior performance of modality-level fusion over decision-level fusion underscores the benefit of deeper multimodal integration in emotion recognition. The \textbf{EMO-MSASE} framework not only advances the capabilities of affective computing but also provides a robust, scalable approach for applications in adaptive virtual environments, and human-computer interaction. These results highlight the potential of \textbf{EMO-MSASE} to set a new benchmark in multimodal emotion recognition, paving the way for further exploration and innovation in multi-domain deep learning architectures.

\section{Acknowledgments}
The project was approved by the Deakin University Human Ethics Advisory Group - Faculty of Science Engineering and Built Environment (SEBE-2023-20).

\bibliographystyle{IEEEtran}

\bibliography{rf}

\vfill

\end{document}